\documentstyle[psfig,manuscript,aps]{revtex}
\tightenlines
\begin{document}
\title{On the effect of short-range magnetic ordering on electron  
energy-loss spectra in spinels}

\author{D.W. McComb \footnote{Now at department of Materials, 
Imperial College LOndon, London SW7 2AZ. Author to whom correspondence
should be addressed: dcomb@imperial.ac.uk} \\
Department of Chemistry, University of Glasgow, Glasgow G12 8QQ }

\author{A.J. Craven \\
Department of Physics and Astronomy, University of Glasgow, Glasgow G12 8QQ}

\author{L. Chioncel and A. I. Lichtenstein  \\
Institute for Theoretical Physics, University of Nijmegen, 
6500 GL Nijmegen, Netherlands }

\author{F.T. Docherty  \footnote{Now at Department of Pure and Applied
Chemistry, Strathclyde University, Glasgow}\\
Department of Chemistry, University of Glasgow, Glasgow G12 8QQ \\
Department of Physics and Astronomy, University of Glasgow, Glasgow G12 8QQ}

\maketitle
\begin{abstract}
The energy-loss near-edge structure (ELNES) at the oxygen $K$ edge in two $AB_2O_4$ 
spinels (A=Mg, B=Al,Cr) is reported.  In $MgAl_2O_4$ the experimental data is 
successfully modelled within the LDA framework. In the case of $MgCr_2O_4$ 
spin-polarisation, in the form of antiferromagnetic ordering on the Cr 
sub-lattice, must be included despite the fact that the measurements were 
performed at approximately 30 times above $T_N$.  A model in which dynamic 
short-range antiferromagnetic ordering is present at room temperature is 
proposed to explain the results of the experiments and calculations.      
\end{abstract}

\section{Introduction}

Electron energy-loss spectroscopy (EELS) carried out in the nano-analytical 
electron microscope (nano-AEM) is a powerful probe of chemistry, bonding and 
electronic structure in a wide range of materials. Analysis of energy-loss 
near-edge structure (ELNES) present on ionisation edges in EELS is now 
recognised as the only technique that can be used to obtain such information 
with close to atomic-scale spatial resolution \cite{Egerton96}. The ELNES technique can be 
used to obtain information on coordination number, local symmetry and oxidation 
state to investigate chemical and structural inhomogeneities in advanced 
materials e.g. in the vicinity of interfaces, grain boundaries etc. In 
combination with other microscopy techniques such as electron diffraction and 
high resolution electron microscopy, ELNES can be utilised to obtain complete 
characterisation across interfaces. However the electron-specimen interactions 
that result in ELNES are complex.  In order to advance the use of the technique 
it is important to conduct studies that combine fundamental investigations of 
the ELNES in known structures and theoretical modelling of the electron-specimen 
interactions. 

We have previously reported the results of a combined experimental and theoretical 
study on a series of ternary transition metal oxides that exhibit the spinel 
structure \cite{Docherty01}.   
The oxygen K-edge ELNES from a range of $AB_2O_4$ materials (A=Mg, Ni, Zn; B=Al, Cr) 
were presented and the results were analysed with the aid of the results of 
electronic structure calculations performed using a multiple scattering algorithm.  
While good agreement between experimental and theoretical spectra was obtained for 
$MgAl_2O_4$, it was extremely poor when the Al$^{3+}$ cations were replaced by Cr$^{3+}$.   
$MgCr_2O_4$ undergoes antiferromagnetic ordering with a N\'eel temperature 
($T_N$) of 16K \cite{Shaked70}.    
An earlier study of transition metal carbides and nitrides with the rock salt 
structure included data from CrN, which has a N\'eel temperature of 288K.     
The shape of the experimental N K-edge ELNES from CrN does not change over the 
temperature range from 133K to 433K so that long-range magnetic order has no effect.   
Moreover, the x-ray absorption near-edge structure (XANES) matches the ELNES 
extremely well demonstrating that the dipole approximation holds well under the 
experimental conditions used to obtain the ELNES.   With the exception of CrN, 
the agreement between experiment and theory, assuming random spins, is excellent.   
However, while the theoretical shape for CrN follows the pattern observed in the 
related compounds, the experimental shape from CrN differs significantly, 
indicating that some other effect is present.   The agreement between experiment 
and theory for CrN is improved if the spins in the calculation are polarised to 
match the antiferromagnetic structure determined by Corliss et al. \cite{Corliss60}.
   The reason 
for this is that the decrease in energy on going from the unpolarised to the 
polarised state is 336meV per formula unit so that the thermal energy available 
at 433K is unlikely to destroy the polarisation over the short range even though 
the long range order disappears. This observation led us to explore the 
effect of the antiferromagnetic coupling between the chromium sites on the 
calculated ELNES to see if the agreement with experiment was improved despite the 
fact that the measurements were performed well above the N\'eel temperature.

\section{Materials and Methods}

The spinel specimens were synthesised from the parent binary oxides using 
standard solid-state procedures.  The lattice parameters, oxygen parameter, $u$, 
and degree of inversion, $\lambda$, were obtained by Rietveld refinement of X-ray or 
neutron diffraction data.  The $MgCr_2O_4$ specimen was identified as cubic spinel, 
$a=8.33202(1) \AA$, with an oxygen parameter, $u$, of $0.26124$ which is close to the 
ideal value of $0.25$.  Rietveld refinement of the site occupancies revealed the sample 
to be $100 \%$ normal ($\lambda=0$). $MgAl_2O_4$ was also identified as cubic spinel, 
$a=8.08098(2) \AA$, with an oxygen parameter close to the ideal value ($u=0.26116$).  
Rietveld refinement of the site occupancies revealed the sample to be partially 
inverse ($\lambda=0.32$). Samples were prepared for EELS by crushing the powder and then 
dispersing the particles in propan-2-ol.  A few drops of the suspension were then 
placed on holey carbon film coated grid and the solvent was allowed to evaporate.  
The EELS data were collected using a VG HB5 cold field emission gun (FEG) scanning 
transmission electron microscope (STEM), operated at 100 kV.  The spectra were 
recorded using a Gatan 666 parallel electron energy-loss spectrometer.  A collection 
half-angle of 12 mrad was used, defined by a 1mm aperture preceding the STEM 
detectors, and a spray aperture was positioned after the same detectors to remove 
any stray scattering from the collector aperture.  Spectra were recorded at an 
energy-dispersion of 0.1eV/channel, using a 1 nm electron probe containing $~ 0.2$ nA 
with a convergence half-angle of 11mrad.   

The ab-initio electronic structure calculations were carried out using the scalar 
relativistic linear muffin-tin orbital (LMTO) method within the atomic-sphere 
approximation in two flavours: LDA and LDA+U \cite{Andersen,Anisimov}. The MT-radii used in all calculations 
were 2.32, 2.44 and 1.95 a.u. for Mg, Cr and O, respectively. Two additional classes of 
empty spheres were introduced with $R=$1.82 a.u. and $R=$1.66 a.u. respectively. The frozen 
core approximation was employed in the generation of the potential and the parametrization 
of von Barth-Hedin \cite{Barth72} was used for the exchange 
correlation part of the effective 
one-electron potential, obtained within the local spin density approximation of the 
density functional theory. Convergence of charge density was achieved within $10^{-5}$. 
The integration over the Brillouin zone has been done by the tetrahedron method with 
64 k-points for the 
Self-consistent field calculation.  Although all of the Cr atoms in the normal 
spinel structure are crystallographically equivalent, in the spin-polarised calculation 
they were treated differently because of the antiferromagnetic coupling. In the spinel 
structure, the Cr ions form a corner shared tetrahedral lattice, comprising the $16d$ 
sites in space group $Fd3m$, that is normally referred to as the pyrochlore lattice.  
In the calculations, an antiferromagnetic structure in which two of Cr atoms among four 
Cr atoms in each tetrahedron of the pyrochlore lattice are in one orientation of 
magnetization while the other two Cr atoms are aligned in the opposite orientation. 
Although the actual magnetic structure of $MgCr_2O_4$ is likely to be more complex than 
this simplified model, this antiferromagnetic structure was taken as the initial 
state of the calculation and was preserved after the self-consistent calculation.

\section{Results and Discussion}

The intensity observed in an electron energy-loss spectrum is related to the 
product of an atomic transition matrix element and the symmetry-projected unoccupied 
DOS on the atomic site of interest.  Over the limited energy range of interest for 
analysis of fine structure on an ionisation edge, typically $<15$eV, it is reasonable 
to assume that the transition matrix element is a slowly varying function.  Thus, 
to a first approximation, it is reasonable to compare the observed ELNES with the 
broadened site- and symmetry-projected unoccupied DOS obtained from the theoretical 
calculations \cite{Egerton96}.  The calculated DOS is convoluted with two 
functions; a Lorentzian 
function whose width varies with energy above the threshold to simulate the broadening 
from the finite lifetime of the excited state and a Gaussian (FWHM=0.8 eV) to account 
for the experimental resolution at the energy-loss of the oxygen K-edge.  The lifetime 
broadening is approximated by two contributions.   One is from the lifetime of the 
core hole, which can be obtained from atomic data (0.11eV for oxygen 1s) \cite{Stohr96}, 
and the 
other is from the lifetime of the excited atomic electron.  The latter can be 
estimated using the expression of Muller et al assuming a valence band width of 
18eV and a plasmon energy of 25eV \cite{Muller98}. 

In Figure 1 the experimental oxygen K-ELNES of $MgAl_2O_4$ is compared with the result 
of the LDA calculation.  In this calculation the oxygen K-edge was simulated assuming 
that the spinel structure was $100 \%$ normal.  Although experimental data was recorded from 
a partially inverse material, it was concluded in our earlier study that the effect of 
inversion on the oxygen K-edge was minimal and hence this was a valid assumption 
\cite{Docherty01}.
This conclusion is supported by the excellent agreement between experiment and theory 
demonstrated in Figure 1.  The result of the LDA calculation is consistent with the 
calculated band structure results reported by other workers using different algorithms.   

When the octahedral site cation is changed from Al$^{3+}$ to Cr$^{3+}$ the agreement between 
experiment and theory is much poorer (Figure 1).   The results of the neutron 
diffraction studies reported previously clearly demonstrate that the $MgCr_2O_4$ specimen 
is $100 \%$ normal\cite{Docherty01}.
  This is as expected since placing the Cr$^{3+}$ ion in a tetrahedral 
environment would be energetically unfavourable due to the large Jahn-Teller distortion 
that would result from unsymmetric occupation of the frontier orbitals \cite{Cox95}. 
 This is 
significant since it establishes that the discrepancy between experiment and theory 
is not associated with structural inversion.

In the initial LDA calculations (Fig. 1), it was assumed that the band structure of 
the material is independent of electron spin.   However, it is well established that 
this is not necessarily true in a transition metal system where spin-ordering on the 
metal sites could influence the band structure through the exchange-interaction term.  
It is also possible that the low dispersion of the d-band in the vicinity of the Fermi 
level may induce strong electron correlation effects and thus invalidate the independent 
electron model.  The former hypothesis could be investigated by introducing spin 
polarisation into the calculation while the latter requires incorporation of electron 
correlation effects via the LDA+U approach.  

The calculated symmetry-projected density of states (DOS) on the anion and cation 
sites are compared in Figure 2.  The upper panel shows the oxygen $p-$ and the chromium 
$d-$DOS obtained from the LDA calculation performed with no spin polarisation.  In this 
non-magnetic structure it is clear that there is a large DOS at the Fermi level ($E_F$) 
and there is no distinct energy gap ($E_g$).  This is a classic situation where the total 
energy of this "metallic" structure could be lowered by spin polarisation.  
In the LSDA calculation, a structure that exhibits antiferromagnetic ordering of the 
electron spins on the chromium sites has been assumed and the O $p-$ and 
Cr $d-$ DOS obtained 
are shown in the middle panel of Figure 2.  Analysis of the Cr $d-$ DOS shows that the 
energy of one of the spin states ($\uparrow$) has shifted to lower 
energy while the other ($\downarrow$) 
has shifted to higher energy. This has resulted in the creation of an energy gap of 
0.7eV between the valence and conduction bands. The moments on the two Cr sites were 
found to be 2.69 $\mu_B$ and -2.69 $\mu_B$ in this antiferromagnetic insulator and 
as expected 
the total energy was found to be significantly less, 0.811 eV/Cr, than the non-magnetic 
structure.  It should be noted that if the Cr cations are antiferromagnetically ordered 
then two magnetically inequivalent oxygen sites must also be present in the structure.  
In a normal spinel such as $MgCr_2O_4$, each O site has four nearest neighbours 
(1 Mg $\&$ 3 Cr) 
in a tetrahedral arrangement.  Consequently in the magnetic structure $50\%$ of the 
oxygens will have two Cr neighbours with parallel and one with antiparallel spin 
alignments and the other half of the oxygens will have the opposite arrangement.  
This means the $p-$ DOS on each oxygen site is slightly different as shown in Figure 2.  
When the AF ordering on the chromium sub-lattice is reversed the $p-$ DOS on each type 
of oxygen site will also change.  However, the symmetry of the lattice and perfect 
AF ordering means that the effect here is simply a switching of the $p-$ DOS between 
the two types of oxygen site. 

In the "LSDA+U" a value of  U=2eV has been used and the same antiferromagnetic 
ordering of the Cr sites, as used in the LSDA calculation, has been assumed.  The 
general shape and interpretation of the site and symmetry projected DOS predicted by 
the LSDA+U calculation is very similar to that obtained from the LSDA calculation.  
(Fig. 2)  The main difference is that the magnitude of the energy gap has now increased 
to 1.3eV, an effect that is expected since the main role of the Hubbard U is to 
stabilise the magnetic structure. In the "LSDA+U" results, the magnetic moments 
on the Cr sites are similar to those found in the LSDA calculation and lattice energy 
was 1.476eV/Cr less than the non-magnetic lattice energy obtained from a LDA+U 
calculation.  Once again the larger stabilisation energy in the LSDA+U calculation can 
be attributed directly to the effect of U.  Nevertheless it is clear that the 
antiferromagnetic state is more favourable than non-magnetic structure for the 
experimental lattice parameters used in the calculation.

Since there are two types of oxygen site in the magnetic structures it was 
necessary to calculate the average density of states by summing the $p-$ DOS for each 
spin state.  Both types of oxygen site have the same average $p-$ DOS.  The $p-$ DOS 
obtained from each theoretical model was broadened as described earlier and compared 
with the experimental ELNES data. (Fig. 3)  It is immediately apparent that changing 
from the non-magnetic structure in the LDA calculation to the antiferromagnetically 
ordered structure in the LSDA calculation results in a dramatic improvement in the 
agreement between theory and experiment.  Further refinement of the calculation to 
include electron correlation effects has only a minor (detrimental) effect on the 
observed shape.      

The implication of these results is that the oxygen K-ELNES (and hence the electronic 
structure) of $MgCr_2O_4$ can be successfully modelled as a magnetically ordered insulator.  
It is known that $MgCr_2O_4$ exhibits antiferromagnetic behaviour below its N\'eel 
temperature of 16K.  However, the ELNES measurements were conducted at room temperature 
where all long-range antiferromagnetic order has been lost.  The only possible explanation 
of these observations is that even in the room temperature paramagnetic state there 
exists short-range (on the order of 1-2 unit cell lengths) antiferromagnetic ordering 
of the spin directions on the chromium sites over a period of time longer than the 
interaction time between this region and the fast electron.   Since the electron 
velocity is $\simeq $10$^8$ m sec$^{-1}$ and the dimensions of the region 
are $\simeq $1nm, the interaction time is $\simeq $10$^{-17}$ sec.

Thus, in order to explain the experimental and theoretical results obtained we are 
proposing that (a) short range $B-B$ site antiferromagnetic correlations exist in 
$MgCr_2O_4$ 
at room temperature, and (b) these correlations can be detected by electron energy-loss 
spectroscopy.  Let us consider each of these hypotheses in more detail.  The fact that 
$MgCr_2O_4$ shows antiferromagnetic behaviour at all demonstrates that the crystal can 
reduce its lattice energy by ordering the electron spins on the $B-$sites.  In the absence 
of any radical change in crystal structure or lattice parameter it follows that 
correlation of the electron spins will lower the lattice energy of the solid at every 
temperature below the melting point.  However, at temperatures above $T_N$, the thermal 
energy will act to constantly randomise the long-range order and average out any 
correlations.  However, given a difference in energy of 811meV/Cr, it would be expected 
that spin correlations would persist at temperatures as high as 9$\times$10$^3$K.
   Consequently 
the important factor is the length of time over which the correlations exist - this 
will depend on temperature, decreasing as the temperature increases.  The distance over 
which the correlations exist will also depend on temperature - for a fixed time 
interval the spatial extent of the ordering will decrease as the temperature increases.

To our knowledge, the existence of short-range order (SRO) in $MgCr_2O_4$ has not been 
investigated in detail previously.  However, SRO in zinc ferrite, $ZnFe_2O_4$, has been 
investigated by several research groups.  The comparison between $ZnFe_2O_4$ and $MgCr_2O_4$ 
is an excellent one since both materials can be synthesised as $100 \%$ normal cubic 
spinels.  
$ZnFe_2O_4$ is also an antiferromagnetic insulator, with high spin 
Fe ($3d^5$), and a low ordering 
temperature ($T_N$ = 10.5 K) \cite{Konig70}. The spin structure of 
the material as a function of 
temperature is complex with dynamic SRO reported to persist up at least 10$T_N$.  
This conclusion was reached on the basis of neutron scattering 
\cite{Schiessl,Chukalkin88},  
Mossbauer resonance \cite{Schiessl,Oliver99},
and muon-spin-rotation/relaxation ($\mu-$SR) (Ref. \cite{Schiessl}) studies. 
Below $T_N$, long range 
order (LRO) has been observed in $ZnFe_2O_4$ by a number of research groups.  However, at 
temperatures up to $\simeq$10$T_N$, broad maxima have been 
observed and a model in which small 
antiferromagnetic regions exhibit "superantiferromagnetic" behaviour due to SRO 
has been proposed \cite{Chukalkin88}. It has been deduced from the width of the 
maxima that the size 
of the domains exhibiting SRO is of the order of 3nm \cite{Schiessl}.  
The results of the $\mu-$SR 
studies clearly demonstrate that the SRO is dynamic rather than static, a conclusion 
supported by recent single crystal neutron scattering studies \cite{Kamazawa99},   
with an estimated 
minimum fluctuation rate of 1.5 GHz, although the error on this estimate could be as 
large as the value itself \cite{Schiessl}.  
This rate corresponds to a correlation time ($\tau_c$) of 
less than 6.6$\times$10$^{-10}$s.  Since the SRO is detectable by neutron scattering, it is 
also possible to derive an upper value for the correlation time from the interaction 
time between a neutron and a domain of this size.  Assuming an energy range 0.1 - 1eV 
we obtain neutron velocities in the range 0.45-1.4$\times$10$^4$ ms$^{-1}$.  
Using the estimated 
domain size from previous studies of 3nm, we conclude that in order for SRO to be detected 
by neutron scattering the correlation time must be greater than 3$\times$10$^{-13}$s.  
This is 
four orders of magnitude greater than the interaction time for 10$^5$eV electrons and so 
it can be concluded that electron energy-loss spectroscopy is potentially sensitive to 
much shorter lived correlations.

The final question that must be considered is the length scale over which the correlation 
must extend in order for the domain to appear "infinite" as far as the electron 
(or X-ray) excitation process is concerned.  This is readily deduced from the multiple 
scattering theory developed to describe the X-ray absorption near-edge structure (XANES), 
which is the X-ray equivalent of ELNES.   This real space multiple scattering approach 
gives the same result as a band structure calculation in reciprocal space provided that 
both calculations converge \cite{Ankudinov98}. The real space multiple 
scattering approach to calculation 
of electronic structure has been shown to be robust and reliable provided a large enough 
cluster is used to ensure convergence.  Based on our own work on spinels, we have 
demonstrated that convergence in $MgAl_2O_4$ is achieved, at the energy resolution available 
in ELNES or XANES, with a cluster radius of less than 0.7 nm \cite{Docherty01}.  
Thus the correlation of 
the spin has only to exist over a distance of $\simeq$2nm for a time $\simeq$10$^{-17}$ sec 
to have a large 
effect on the edge shape in ELNES or XANES. 

In summary, we have demonstrated that the oxygen K-edge ELNES of $MgCr_2O_4$ cannot be 
simulated using electronic structure calculations performed within the LDA.  However, 
reasonable agreement between experiment and theory is achieved when the calculation is 
performed within the LSDA framework and full antiferromagnetic ordering on the Cr sites 
is included in the model.  While the magnetic structure of $MgCr_2O_4$ is likely to be more 
complex than the model assumed, the improved agreement between theory and experiment 
obtained using this model shows that spin polarisation plays a key role in defining 
the electronic structure as determined by EELS.  While studies of the magnetic behaviour 
of $MgCr_2O_4$ above $T_N$ have not been reported in the literature, 
the work performed by other 
groups on $ZnFe_2O_4$, also an antiferromagnetic (AF) insulator, clearly demonstrate the 
presence of short range AF ordering well above $T_N$.  We conclude that electron energy-loss 
spectroscopy is a highly sensitive probe of such ordering and can potentially reveal 
the presence of AF order over distances of $\simeq$2nm.

\begin{figure}
\centerline{\psfig{file=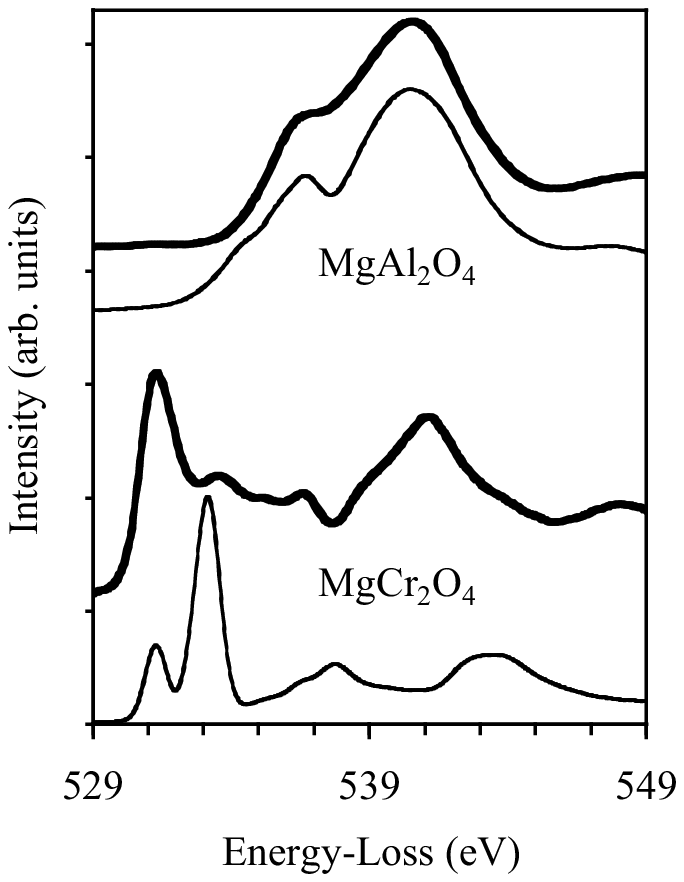,height=3.0in}}
\vskip  0.25cm
\caption{Comparison of experimental oxygen K-ELNES data (thick lines) with the
broadened p-projected unoccupied DOS on the oxygen site obtained from LDA
calculations (thin lines) for $MgAl_2O_4$ and $MgCr_2O_4$. In each case the first peak in
the theoretical result has been aligned to the first peak in the experimental spectrum.}
\label{Fig1}
\end{figure}

\begin{figure}
\centerline{\psfig{file=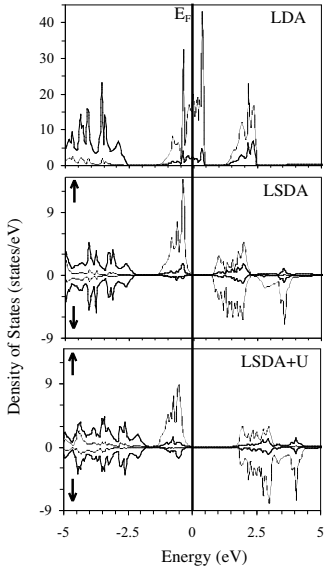,height=3.0in}}
\vskip  0.25cm
\caption{Comparison of the calculated oxygen p-DOS (thick line) and chromium d-DOS 
(thin line) in $MgCr_2O_4$ obtained from different theoretical models. The Fermi energy is 
at 0eV.  The upper panel shows the result for the non-magnetic structure.  In the LSDA 
(middle panel) and LSDA+U (lower panel) calculations, antiferromagnetic ordering on the 
Cr sites has been assumed and the result for each spin state on one of the Cr sites is 
shown.  The density of states on the second Cr site is inversely related to the first 
due to the antiferromagnetic ordering.  }
\label{Fig2}
\end{figure}

\begin{figure}
\centerline{\psfig{file=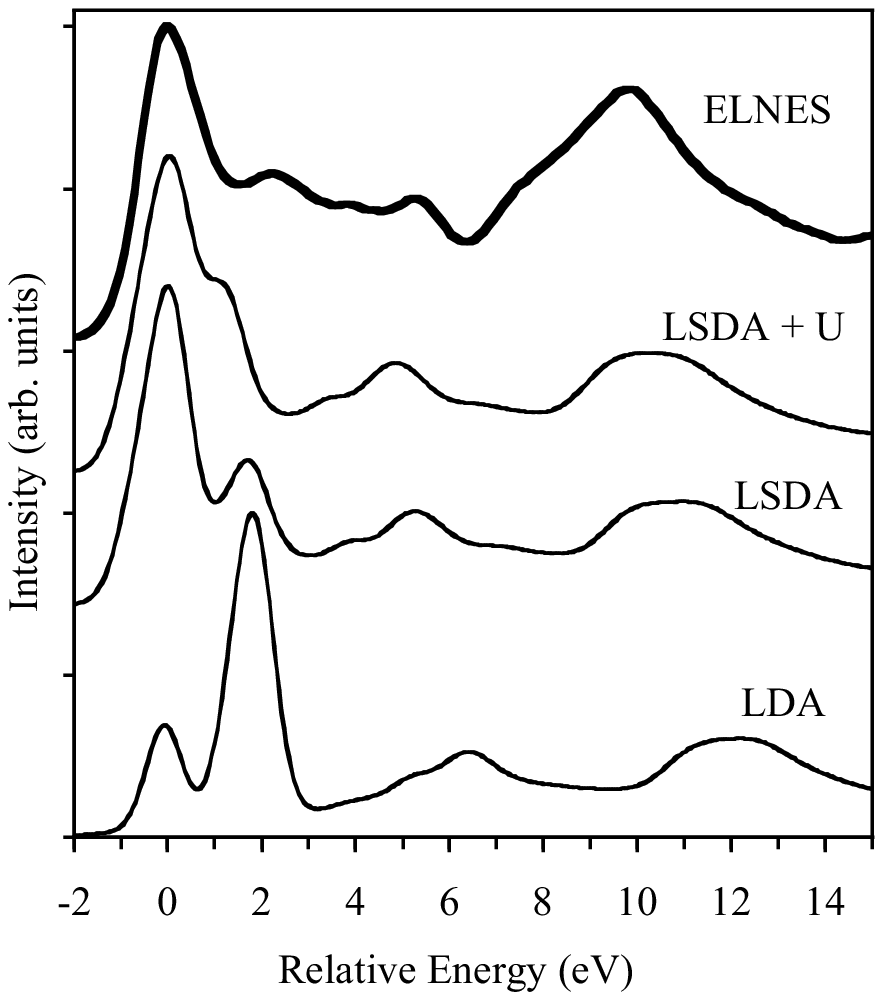,height=3.0in}}
\vskip  0.25cm
\caption{Comparison of the broadened oxygen  p-DOS with the experimental oxygen
K-ELNES data for $MgCr_2O_4$. The first peak in each spectrum has been aligned to 0eV.  
Where antiferromagnetic ordering is present the spectrum is derived from the average 
of the "spin-up" and "spin-down" DOS.}
\label{Fig3}
\end{figure}


\begin{references}
\bibitem{Egerton96}R.F. Egerton, Electron energy-loss spectroscopy in the electron 
microscope, 2nd edition (Plenum, New York, 1996)

\bibitem{Docherty01}F.T. Docherty, A.J. Craven, D.W. McComb, and J.Skakle, 
Ultramicroscopy 86, 273 (2001) 
\bibitem{Shaked70}H. Shaked, J.M. Hastings, and L.M. Corliss,  Phys. Rev. B 1, 3116 (1970). 
\bibitem{Paxton00}A.T. Paxton, M. van Schilfgaarde, M. MacKenzie, and A.J. Craven,  J. Phys.: Condens. Matter 12, 729 (2000). 
\bibitem{Corliss60}L.M. Corliss, N. Elliot and J.M. Hastings Phys. Rev. {\bf 117}, 929 (1960)
\bibitem{Andersen}O.K. Andersen,  Phys. Rev. B 12, 3060 (1975); O.K. Andersen and O Jepsen,
Phys. Rev. Lett. 53, 2571 (1984); O.K. Anderson, O. Jepsen and D. Glotzel in, 
Highlights of Condensed Matter Physics, edited by F. Bassini, F. Fumi and M.P. Tosi 
(North-Holland, New York, 1995); R.O. Jones and O. Gunnarson, Rev. Mod. Phys. 61, 689 
(1989).
\bibitem{Anisimov}V.I. Anisimov, J. Zannen and O.K. Andersen,  Phys. Rev. B 44, 943 (1991); 
V.I. Anisimov, I.V. Solovyev, M.A. Korotin, M.T. Czyzyk and G.A. Sawatzky,  Phys. Rev. B 48,
 16929 (1993); A.I. Lichtenstein, J. Zannen and V.I. Anisimov,  Phys. Rev. B 52, R5467 
(1995); V.I. Anisimov, F. Aryasetiawan, A.I. Lichtenstein,  J. Phys.: Condens. Matter 9, 
767 (1997).
\bibitem{Barth72}U. von Barth and L Hedin, J. Phys. C 5, 1629 (1972).
\bibitem{Stohr96}J. Stohr, NEXAFS Spectroscopy (Springer, New York, 1996)
\bibitem{Muller98}D.A.  Muller, D.J. Singh, and J. Silcox,  Phys. Rev. B 57, 8181 (1998). 
\bibitem{Kostlmeier00}S.  Kostlmeier, and C.  Elsasser, Phys. Rev. B 60, 14025 (1999); 
K. van Benthem and H. Kohl, Micron 31, 347 (2000). 
\bibitem{Cox95}P.A. Cox, Transition Metal Oxides: An Introduction to their Electronic 
Structure and Properties (Oxford University Press, Oxford, 1995).
\bibitem{Konig70}U. Konig, E.F. Bertaut, Y. Gros, M. Mitrikov, and G. Chol, 
Solid State Commun. 8, 759 (1970). 
\bibitem{Schiessl}W. Schiessl, W. Potzel, H. Karzel, M. Steiner, G.M. Kalvius, A. Martin, 
M.K. Krause, I. Halevy, J. Gal, W. Schafer, G. Will, M. Hillberg, and R. Wappling,  
Phys. Rev. B 53, 9143 (1996). 
\bibitem{Chukalkin88}Yu.G. Chukalkin and V.R. Shtirts,  Sov. Phys. 
Solid State 30, 1683 (1988). 
\bibitem{Oliver99}S.A. Oliver, H.H. Hamdeh, and J.C. Ho,  Phys. Rev. B 60, 3400 (1999). 
\bibitem{Kamazawa99}  K. Kamazawa, Y. Tsunoda, K. Odaka, and K. Kohn, J. Phys. 
Chem. Solids 60, 1261 (1999). 
\bibitem{Ankudinov98}A.L. Ankudinov, B. Ravel, J.J. Rehr, and S.D. Conradson,  
Phys. Rev. B 58, 7565 (1998). 
\end{references}
\end{document}